July 24, 2015

# Development of the Pixelated Photon Detector Using Silicon on Insulator Technology for TOF-PET

A.Koyama[1], K.Shimazoe[1], H.Takahashi[1], T. Orita[2], Y.Arai[3],
I.Kurachi[3], T.Miyoshi[3], D.Nio[4], R.Hamasaki[4]

[1] *The Univesity of Tokyo,*
[2] *Japan Atomic Energy Agency,*
[3] *High Energy Accelerator Research Organization,*
[4] *The Graduate University for Advanced Studies*

I describe a feasibility study for development of the pixelated photon detector using silicon on insulator technology

PRESENTED AT

International Workshop on SOI Pixel Detector (SOIPIX 2015)
Sendai, Japan, June 3-6, 2015

# 1    Introduction

  Scintillation detector is widely used for radiation spectrometry. This detector is constructed with scintillator and photo detector. Radioactive rays are converted to visible light by scintillator and photo detector measure this emitted scintillation light (visible light) as electrical signal.

  This scintillation detector has advantage in the point of high detection efficiency compare to other radiation detector, such as semiconductor detector. But scintillation detector has low positional resolution comes from its scintillator's size. If the position resolution of scintillation detector was improved, we would be able to enhance accuracy of radiation spectroscopy.

  Positron emission tomography (PET) spectroscopy is one of the most important clinical applications of scintillation detector. If we achieved development of ultimate positional resolution detector, we will be able to image one molecular and observe functional information. And we are also aiming at using time information between detected opposite gamma rays. This time information is known as time of fright (TOF), and this information can enhance signal to noise ratio of final reconstructed image (Fig.1).

  Therefore, our study motivation is to develop higher performance scintillation detector in the points of time resolution and positional resolution for development of ultimate TOF-PET system.

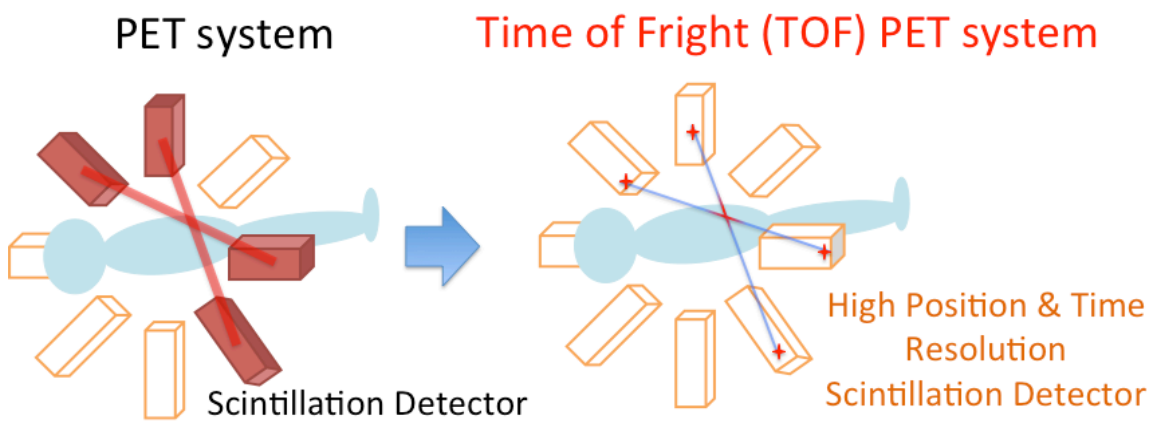

Fig. 1 Schematic of Time of Fright (TOF) – PET system

One solution to improve positional resolution of scintillation detector is to recognize interaction point of gamma ray in scintillator. We are planning to measure emission pattern of scintillation light by using pixelated photo detector and estimate interaction depth in scintillator (Fig.2).

And by using avalanche photo diode (APD) as photo sensor, we will be able to keep higher time responsiveness.

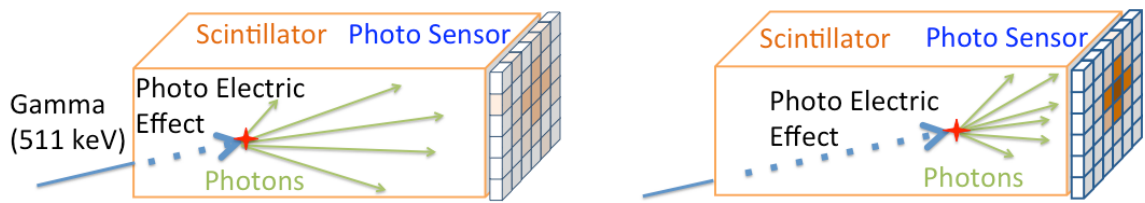

Fig.2 Difference of photon emission pattern in scintillator

# 2 Purpose

To measure light emission pattern in scintillator, higher sensitivity and faster response are required to photo detector. Such as single photon avalanche diode (SPAD), conventional pixelated photo detector is operated at Geiger avalanche multiplication. However higher gain (10000~100000) of SPAD seems very attractive, photon detection efficiency per unit area is low. This weak point is mainly caused by Geiger avalanche mechanism. Each pixel can detect only 1 photon while Geiger avalanche multiplication, and this multiplication requires convergence mechanism by quench circuit.

To overcome these difficulties, we designed Pixelated Linear Avalanche Integration Detector using Silicon on Insulator technology (SOI-Plaid). To avoid dark count noise and dead time comes from quench circuit, we are planning to use APD in linear multiplication mode. This detector has monolithic structure including readout circuit layer and linear APD layer (Fig.3). SOI technology enables laminating each layer, and high-speed and low-noise signal reading regardless smaller gain of linear APD.

This study shows design of linear APD by using SOI fabrication process. We designed test element group (TEG) of linear APD and inspected optimal structure of linear APD.

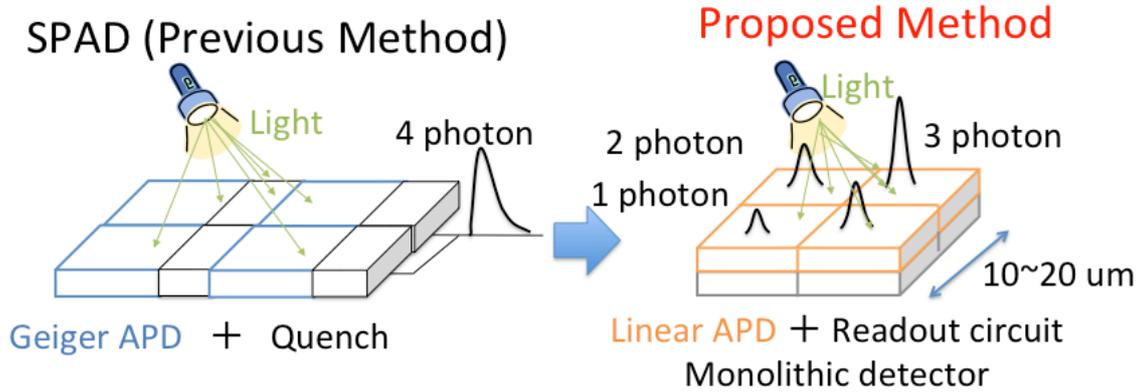

Fig.3　Design of SOI Pixelated Linear Avalanche Integration Detector

# 3　　Method

We fabricated TEG of linear APD on SOI process and simulated electric field distribution of each device. We also performed measurement of current-voltage (IV) characteristics and examined avalanche mechanism.

# 4　　Device fabrication

Figure 4 shows fabricated TEG. We designed 12 structures.　Each pixel size is 20 × 20 μm$^2$ and 400 pixel cells are installed in one array. In this paper, we inspected 4 structures showed at Figure 5.

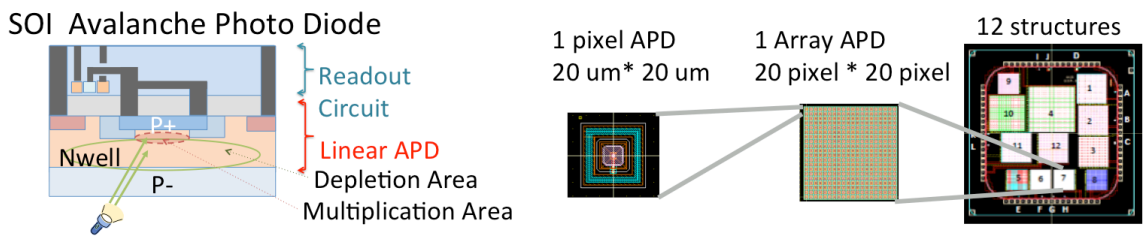

Fig.4 Overview of proposed device (Left) and test element group (Right)

These structures are based on double photo diode method [1]. There are two PN junctions between P+ high dopant area and P-substrate region via N-Well. This structure prevents the deterioration of response properties by removing the accumulation of a slow carriers and collecting avalanched carriers effectively. In this study, we inspected the 4 structures based on double photo diode method. These 4 APDs are different from the points of the depth and the P-Well guard ring. In order to achieve the best performance, APDs require uniformly high electric fields at the p-n planar junction. For this, guard-ring structures, which can relieve the electric field at the edge of p-n junction, are often used [2] [3]. And Deep N-Wells can connect two N-Wells and isolate the guard ring between the N-Wells. Two separate N-Wells can also be connected directly without using deep N-Wells by careful positioning of doping profiles [4]. The internal structure of each element is as follows (Fig.5).

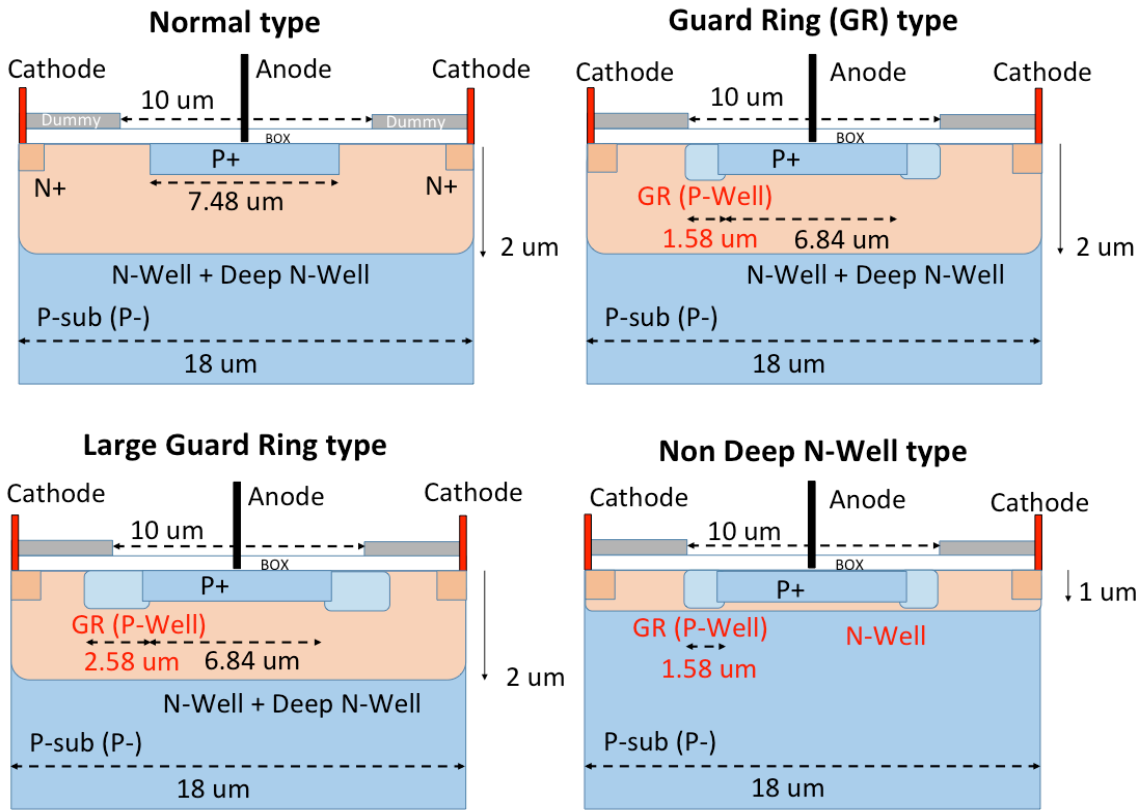

Fig.5 Fabricated APD device structure

# 5  Simulation results

To inspect electric field in fabricated device, we performed simulation by using Technology CAD (TCAD). Figure 6 shows the distribution of electric field at p-n junction edge at reverse bias voltage ($V_{bias}$) = 10[V], 20 [V].

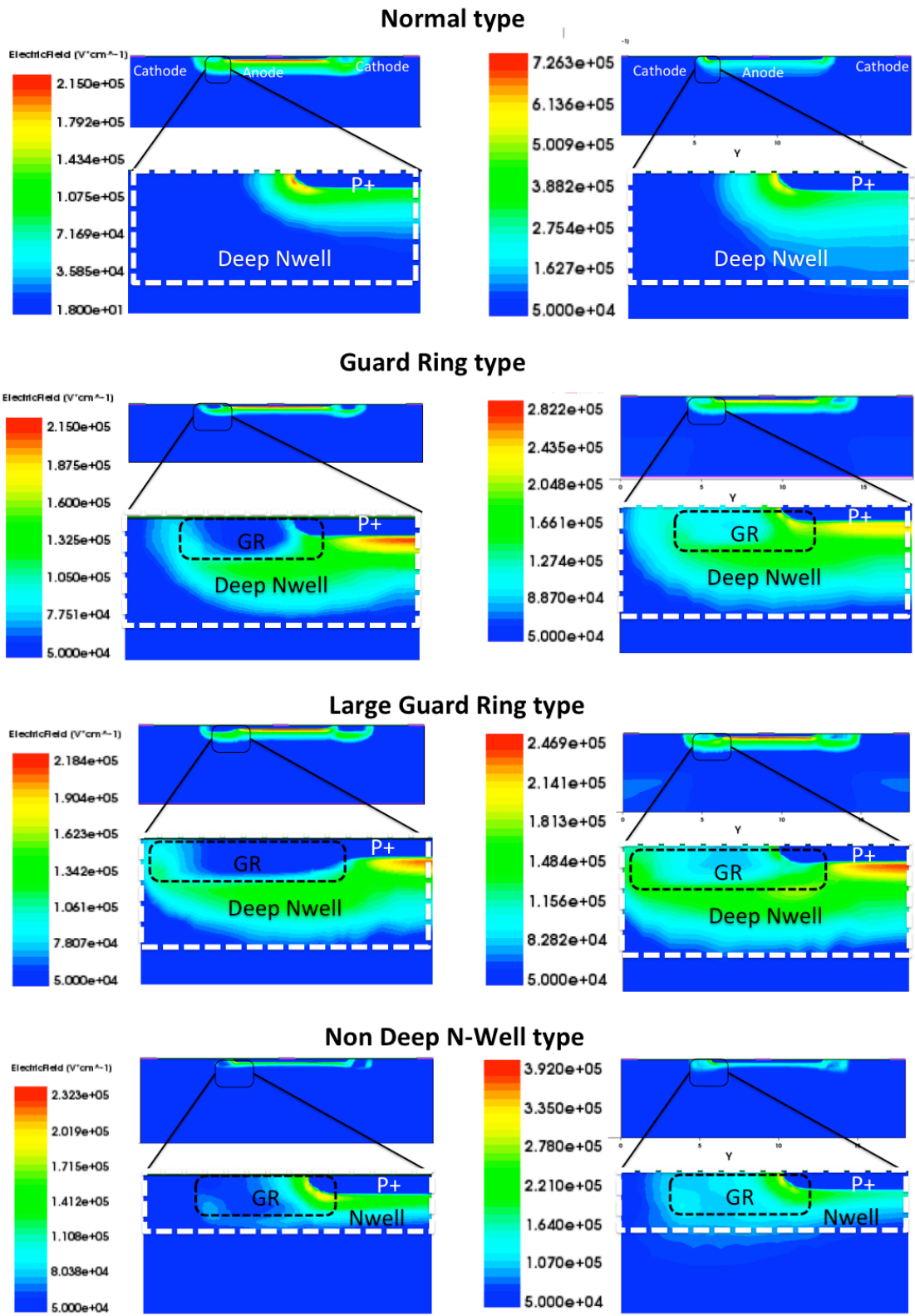

Fig.6 Device simulation results ($V_{bias}$ = 10 [V] (Left), 20 [V] (Right))

From simulation results, Non Deep N-Well type has thinner depletion area and distribution of electric field than other types. We can see high electric field was concentrated at P+ edge and reached 7.2*10^5 [V/cm] at Normal type. On the other hand, Guard Ring type shows high electric field at P+ edge seemed to be controlled effectively by P-Well guard ring.

# 6     Measurement results

We measured current-voltage (IV) characteristics to inspect light response of fabricated APD. In the measurement, we observed the current-voltage characteristics under three different lightning conditions; dark, LED illumination 1.73 [$\mu$W/mm$^2$] and LED illumination 2.9 [$\mu$W/mm$^2$]. LED dominant wavelength was 570 nm. And LED light intensity was measured by using reference photo diode (S1787, HAMAMATSU). All measurements were performed by using semiconductor evaluation analyzer (4200SCS, Keythley).

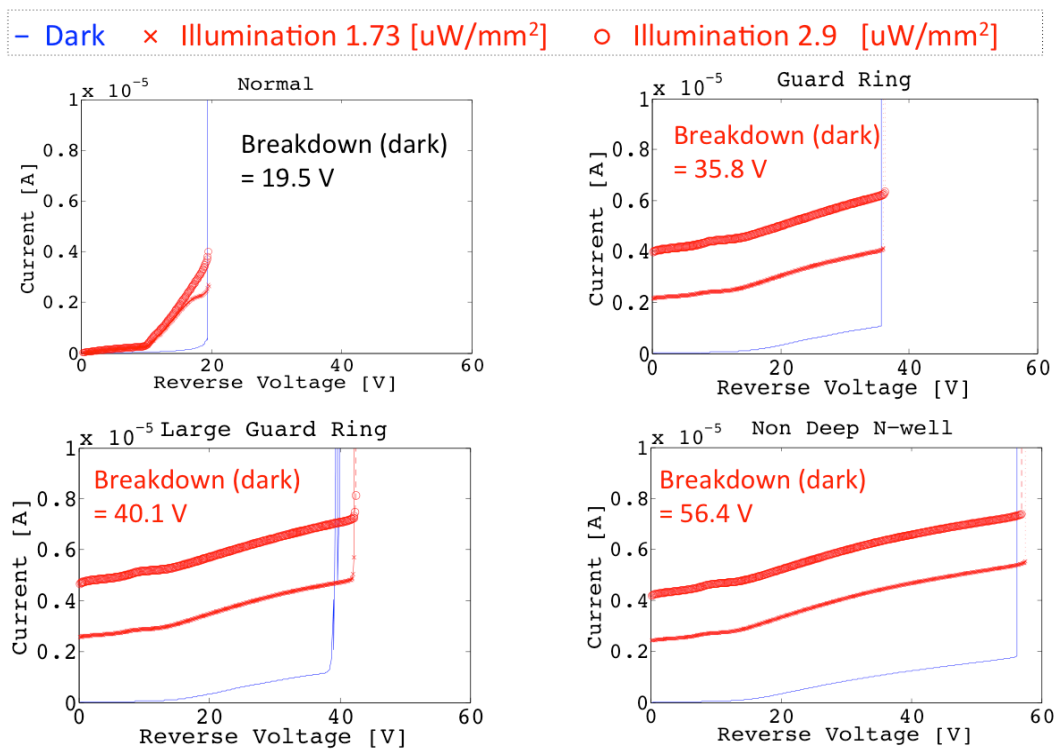

Fig.7    Measurement results of Current-Voltage Characteristics

From Figure 7, Measurement results showed all APD has light response. Guard ring structure was higher breakdown voltage than normal type, which has not guard ring. And Non deep N-Well type showed highest breakdown voltage.

Figure 8 shows calculated gain from IV characteristics. The increase in photo current value was very small value and gain was increased no more than 1.4 by the bias voltage at Large Guard Ring type.

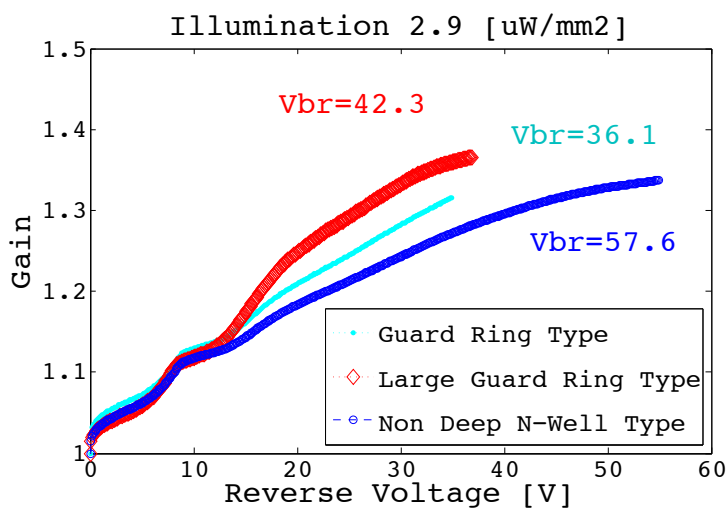

Fig.8 Calculated GAIN of all APDs

# 7 Discussion

Simulation results showed Normal type, which has not guard ring had very high electric field at P+ edge and this reached $7.2*10^5$ [V/cm] at Bias=20 [V].

On the other hand, guard ring structure seems very effective to control concentration of high electric field at P+ edge. Deep N-Well is also effective to enlarge electric field in the depletion area.

Measurement results showed guard ring structure has higher breakdown voltage. This indicated Guard ring structure will be effective to control electric field at edge and makes breakdown voltage higher. And the result of larger guard ring type had higher breakdown voltage. This means the tunneling effect was suppressed more significantly at P+ edge region and it result in higher breakdown voltage.

Measurement results also showed Non Deep N-Well structure has higher breakdown voltage. And simulation results showed vertical spread of electric field was restricted in Non deep N-Well structure. Thus these results indicated electric field was gradually increased in narrow region with bias voltage in Non deep N-Well structure, and this resulted in higher breakdown voltage.

Calculated gain indicated increase of photocurrent would be caused by not avalanche multiplication but improvement of carrier collection efficiency. This suggests that most of the photocurrent was collected in the DNW-P-sub junction and did not contribute avalanche multiplication.

# 8 Conclusion

APD basic property was inspected by fabricated SOI-APD TEG. Measurement results showed almost APD has photocurrent to incident light but it was difficult to observe avalanche multiplication effect. On the other hand, comparison of measurement and simulation results indicated guard ring prevent edge breakdown and deep N-Well is effective to optimize electric field distribution.